\documentclass[runningheads]{llncs}

\usepackage{textcomp}
\usepackage{graphicx}
\usepackage{verbatim}
\usepackage[breaklinks=true]{hyperref}
\usepackage{breakcites}
\usepackage[utf8]{inputenc}
\DeclareUnicodeCharacter{FB01}{fi}
\begin{document}
\title{Measuring the effectiveness of code review comments in GitHub repositories: A machine learning approach
}
%
\author{Shadikur Rahman \and Umme Ayman Koana \and Hasibul Karim Shanto \and Mahmuda Akter \and Chitra Roy \and Aras M.Ismael
}
\authorrunning{  R. Shadikur \& A. Umme Ayman et al}
%
\institute{\{Daffodil International University,  Bangladesh University of Engineering and Technology, Ahsanullah University of Science and Technology\}, Dhaka, Bangladesh\\
Khulna University of Engineering \& Technology, khulna, Bangladesh\\
Sulaimani Polytechnic University, Kurdistan, Iraq \\
\email{shadikur35-988@diu.edu.bd, \{koana2k12, hasibsourov36\}@gmail.com, mahmuda\_akter03@yahoo.com, Chtrroy13@gmail.com, Aras.masoud@usa.com}
}
%
\maketitle              
\begin{abstract}
This paper illustrates an empirical study of working efficiency of machine learning techniques in classifying code review text by semantic meaning. The code review comments from the source control repository in GitHub were extracted for development activity from the existing year for three open-source projects. Apart from that, Programmer\textquotesingle s need to be aware of their code and point out their errors. In that case, it is a must to classify the sentiment polarity of the code review comments for avoiding an error. We manually labeled 13557 code review comments generated by three open source projects in GitHub during the existing year. In order to recognize the sentiment polarity (or sentiment orientation) of code reviews, we use seven machine learning algorithms and compare those results to find the better ones. Among those Linear Support Vector Classifier(SVC) classifier technique achieves higher accuracy than others. This study will help the programmer\textquotesingle s to make any solution based on code reviews by avoiding misconceptions.

\end{abstract}
\section{Introduction}
The code review comments are the key significative of the substance of any coding bug issue or error. By reading any review comments, programmers can understand any coding errors. In most cases, developers do not read the whole code as they think the review comment is the essence of any errors. Sometimes programmers are misguided by reading the review comment on any coding issues. In~\cite{bosu2015characteristics}, it is investigated that how code review comments useful the coder logic and influencing them towards error solving. In that case, it is a must to identify the sentiment polarity of the review comments for avoiding misconception.

Depending on the context of the review comments, the semantic orientation would be different. Moreover, any specific comments sometimes presented differently in different review comments. So it is essential to find the semantic orientation of any code review comment. There are missing enough studies to find out the semantic orientation of a review comment based on the context of coding bug. Open source software measures peer code review impact in~\cite{rigby2012contemporary}, the method of analyzing code written by the different developers on the project to decide whether it is of sufﬁcient quality to be integrated into the project codebase. 

In this study, we present the effectiveness of machine learning and deep learning techniques to find the semantic orientation of any review comments. We use Naive Bayes~\cite{rish2001empirical}, Multinomial Naive Bayes (MNB)~\cite{rennie2003tackling}, Bernoulli Naive Bayes~\cite{mccallum1998comparison}, logistic regression~\cite{ho1994decision}, Stochastic Gradient Descent (SGD)~\cite{bottou2010large}, Linear Support Vector Classifier(SVC)~\cite{gunn1998support}, and Nu support vector classifier~\cite{scholkopf2000new}, and
Word2vec method~\cite{goldberg2014word2vec} to find out the most appropriate semantic orientation. The result of this study will help programmers. 

This paper is structured as follows: Related work is described at~\ref{Related Work} that followed by Research Methodology and Result \& Discussion at~\ref{Research Methodology} and~\ref{Result & Discussion} respectively. The final section~\ref{Conclusion} summarizes our contribution and furnishes the conclusion.
\section{Related Work}\label{Related Work}
The sentiment analysis of code review comments from GitHub projects repository consists of a big stages: the sentiment analysis of all the code review comments.It is shown~\cite{pletea2014security} that the sentiment analysis of the detection of comments related to the security topic and the sentiment analysis of all the comments from GitHub projects. Many works have been done on product reviews using opinion mining and sentiment analysis. In~\cite{rahman2019context} that most of the sentiment analysis was done on news headlines data. Other than that news data, social media data and web blogs data are also used.  In~\cite{pang2002thumbs}, Naive Bayes classification, maximum entropy classification, and support vector machines these tree machine learning algorithm were used to perform sentiment analysis on movie review. 
A feature-based opinion summarizing technique is presented in~\cite{hu2004mining} of product reviews using data mining and natural language processing. A system Opinion Observer~\cite{liu2005opinion} is implemented using proposed holistic lexicon-based approach~\cite{ding2008holistic}. A lexicon-based approach is used in~\cite{im2013analysing} to identify the positive or negative polarity of the financial news. A large-scale sentiment analysis system is presented in this paper~\cite{godbole2007large} to indicate positive and negative opinion of news and blogs by assigning score. In~\cite{maynard2012challenges} the author performed a sentiment analysis by indicating positive, negative or neutral sentiments. Sentiment Analysis tools are based on SentiCR In~\cite{ahmed2017senticr} of social media text or product reviews. Opinion Lexicon-based algorithm and Naive Bayes algorithm are used in~\cite{shuhidan2018sentiment} for sentiment analysis of financial news headlines of Malaysia.

In recent, deep learning is also used for sentiment analysis. Analyzing emotion hidden of the microblog messages in~\cite{xue2014study} Sentiment Dictionary using Word2vec tool based on our Semantic Orientation.  In~\cite{severyn2015twitter} uses deep learning approach for predicting polarities of tweets at both message level and phrase level. Micro blogging and movie reviews datasets are used in~\cite{araque2017enhancing} to measure the performance of sentiment analysis using deep learning. In~\cite{tang2015deep}, authors provide an overview of deep learning approaches for sentiment analysis and also suggest some mitigation to address the challenges.

\section{Research Methodology}\label{Research Methodology}
\begin{figure}
\includegraphics[width=\textwidth]{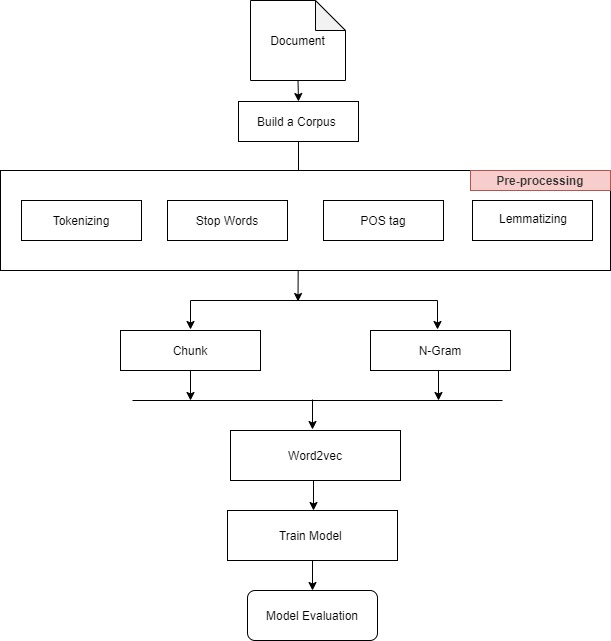}
\caption{Research Methodology} \label{research_design}
\end{figure}

In this section, the entire process of our research activities has been described. First of all, we have chosen our  dataset\footnote{https://github.com/sadirahman/Code-review-sentiment-analysis/tree/master/code-reviews}. For the dataset, we have selected code review comments to perform our research process. For preparing so, we have a lot of covers to process e.g. step by step preprocessing, chunking, N-gram, training model with the help of machine learning and deep learning methods. We convey a retrieve responsibility to compare three topic representations: (1) Sentiment analysis (2) Word scoring and (3) Measuring comments. Figure \ref{research_design} shows an overview of our research experiment.
\subsection{Training Dataset Generation}
We utilized according to the three-step approach to build a labeled sentiment dataset from code review comments. We focused only on the structure code review comments. Table \ref{Dataset} summarizes the statistics about these three projects data sets.

\paragraph{\textbf{GitHub Api:}}
We used GitHub api\footnote{https://api.github.com/} to mine the code review comments of our three popular projects.We used the repos-url to fetch all repositories of our projects. We had worked on several languages including HTML, CSS, C++, Java, Python, and others. Our projects have performed at least 5,000 code reviews.

\begin{table}[hbt!]
\centering
\setlength{\tabcolsep}{10pt}
\caption{Statistics of Data sets}\label{Dataset}
\begin{tabular}{l c r r}
\hline
Project Name & Time length & Comments &  Ratio\\
\hline
Alpha  &  Jan, 2019 to Nov, 2019 & 4867 & 35.90\%\\
Beta  &  Jan, 2019 to Nov, 2019 & 5655 & 41.71\%\\
Gamma  & Jan, 2019 to Nov, 2019 & 3035 & 22.38\%\\\hline
& Total & 13557 & 100\%\\
\hline
\end{tabular}
\end{table}
\subsection{Data Categorizing}
After collecting data sets, we categorize the comments based on structure reviews. For this paper, we mainly consider only 4 categories. For doing this, we categorize the comments using the meta data of each project\textquotesingle s URL. Table \ref{Cetagories} shows the categories of the reviews of we identified. The distribution of the number of reviews items for each project is shown in Table \ref{Distribution}.
\begin{table}[hbt!]
\centering
\setlength{\tabcolsep}{8pt}
\caption{Categories of code review comments}\label{Cetagories}
\begin{tabular}{l l}
\hline
Labels & Comments \\
\hline
Efficient  &  Please reformat in accordance to previous class\\
Not-Efficient  & Sure sounds good \\
Some-How-Efficient  & I think this information is obsolete now \\
System-Generated  & Change has been successfully merged by Sadi Rahman \\

\hline
\end{tabular}
\end{table}

\begin{table}[hbt!]
\centering
\setlength{\tabcolsep}{10pt}
\caption{Distribution of review comments based on labels}\label{Distribution}
\begin{tabular}{llll}
 \hline
 Labels & Alpha  & Beta  & Gamma  \\ 
 \hline
Efficient & 44\%    & 31\%  &   29\% \\
Some-How-Efficient & 30\%   & 19\%   & 23\%  \\
Not-Efficient & 11\%  & 14\%  & 13\%   \\
System-Generated & 15\%  & 36\%  &  35\%    \\

 \hline
\end{tabular}
\end{table}
\subsection{Data Preprossessing}
\begin{figure}
\includegraphics[width=\textwidth]{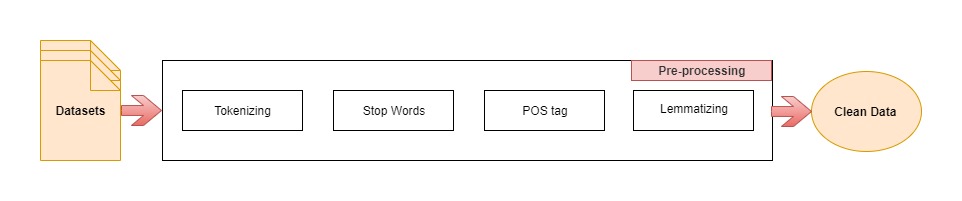}
\caption{Data Preprossessing process} \label{Preprossessing}
\end{figure}

To analyze our data set, we preprocess all the review comments. In most of the cases, we realize that text data is not perfectly cleaned. For cleaning the text data, text pre-processing is needed.  For pre-processing,  we  need  to  follow  several  steps  like tokenization, stop words removing, POS tag, lemmatizing and removing punctuation. Figure \ref{Preprossessing} shows an overview of our data prepossessing process.
\paragraph{\textbf{Text Tokenization:}}
Tokenization is the method of splitting the provided text into smaller portions called token. Words, numbers, punctuation marks, and others can be recognized as the token.
\paragraph{\textbf{Stop Words Removing:}}
Stop words are the usual common words in a language like “about”, “an”, “the”, “is”, “into”. These words do not give important meaning and are normally removed from text documents of our dataset.
\paragraph{\textbf{Lemmatizing:}}
We used lemmatizing process of decreasing words to their word Lemma, base or root form, for example, roads\textendash road, loved\textendash love.
\paragraph{\textbf{Removing punctuation:}}
Remove punctuation is needed if they are not related to the text corpus. Normally, regular expressions are used to remove set of punctuation symbols.
\paragraph{\textbf{POS tag:}}
We used NLTK word tokenizer to parse each text into a list of words.After Text Tokenization,we used Pos tagging in NLP using NLTK. The part of speech(POS tag) explains corpus how a word is used in a sentence.POS tag is separated into subclasses. POS Tagging solely means labeling words with their appropriate Part-Of-Speech.There are eight main parts of speech - nouns, pronouns, adjectives, verbs, adverbs, prepositions, conjunctions and interjections.Parts of Speech (POS) tag from Penn Tree bank annotation~\cite{taylor2003penn} is shown in Table \ref{POS}.
\begin{table}[hbt!]
\centering
\setlength{\tabcolsep}{10pt}
\caption{POS tagging Penn Tree bank annotation}\label{POS}
\begin{tabular}{lll}
 \hline
 Types of POS  & Initial  & Examples   \\ 
 \hline
Noun & N  & Daniel, table, happiness, hope  \\
Verb & V & go, run, live, like, are \\
Adjective & ADJ  & big, happy, green, young \\
Adverb & ADV  & quietly, very, always, never    \\
Preposition & P &  at, on, in, from    \\
Conjunction & CON & but, because, so, yet   \\
Pronoun & PRO  & you, we, they, he     \\
Interjection & INT  & Ouch!, Wow!, Great!, Help!    \\
 \hline
\end{tabular}
\end{table}

\subsection{Chunking Process}
\begin{figure}
\includegraphics[width=\textwidth]{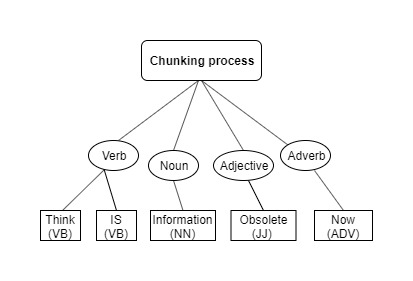}
\caption{Overview of Chunking Process} \label{Chunk}
\end{figure}

After the POS tag, we used the chunking process of extracting phrases from unstructured corpus text.Chunking works on top of POS tagging, it uses pos-tags as input and provides chunks as output.We search for chunks corresponding to an individual some POS tag phrase. The review comments contains many parts of speech which are irrelevant to detect semantic orientation in our case. We consider only Adjectives (JJ), Verb (VB), Adverb (ADV) and Noun (NN) Parts of Speech (POS) tag from Penn Tree bank annotation.For example,code review comment is "I think this information is obsolete now".Figure \ref{Chunk} shows an overview of our chunking process

\subsection{N-Gram Process}
An N-gram is a sequence of N words, which computes p(w|h), the probability  of a word w* \cite{brown1992class}. We have used N-gram model using the same training text and held-out data as we used for the word-based natural language process model we discussed above in our research. The purpose of using this is to maintain the sequence of candidate labels in our training data sets. For example, if we consider the text- "Please enter your comment". For this text, the 2-gram and 3-gram words is presented in Table~\ref{n-gram}. 

\begin{table}[hbt!]
\centering
\setlength{\tabcolsep}{8pt}
\caption{Process of an N-gram words}\label{n-gram}
\begin{tabular}{c c}
\hline
2-gram(bigram) & 3-gram(trigram)  \\
\hline
"Please enter" &"Please enter your"	\\
"Enter your" & "Enter Your comment"\\
"Your comment" \\

\hline
\end{tabular}
\end{table}

\subsection{Word2vec Process}
After the preprocessing, we used the word2vec method in our research work. Word2vec gives direct access to vector representations of training corpus words. Word2vec is to classify the vectors of similar words together in vector space. It recognizes similarities mathematically. Word2vec produces vectors that are distributed numerical representations of word features, features such as the context of individual words.We train Word2Vec Model based on code review corpus and gensim word2vec module. We used word2vec for the most similar word-finding parameter "topn = 10".For example, the word "error" is a vector representation Table~\ref{wordvec}. 
\begin{table}[hbt!]
\centering
\setlength{\tabcolsep}{8pt}
\caption{Process of Word2vec}\label{wordvec}
\begin{tabular}{l c}
\hline
Similar Words(Error) & Numerical Representations  \\
\hline
Holes & 0.93 \\
Necessity & 0.91 \\
Drought & 0.91 \\
Storm & 0.88 \\
Blade & 0.88 \\
Flies & 0.88 \\
Washing & 0.87 \\
Cluster & 0.87 \\
Distresses & 0.87 \\
Lazarus & 0.87 \\
\hline
\end{tabular}
\end{table}

\subsection{Training Model}
In this section, for training our model, we have used seven machine learning algorithms.The used algorithms are:Na\"ive Bayes~\cite{rish2001empirical}, Multinomial Naive Bayes (MNB)~\cite{rennie2003tackling}, Bernoulli Naive Bayes~\cite{mccallum1998comparison}, Logistic Regression~\cite{ho1994decision}, Stochastic Gradient Descent (SGD)~\cite{bottou2010large}, Linear Support Vector Classifier(SVC)~\cite{gunn1998support}, and Nu support vector classifier~\cite{scholkopf2000new}. The data set for training model is available at \footnote{\label{note1}https://github.com/sadirahman/Code-review-sentiment-analysis }.
\begin{figure}
\includegraphics[width=\textwidth]{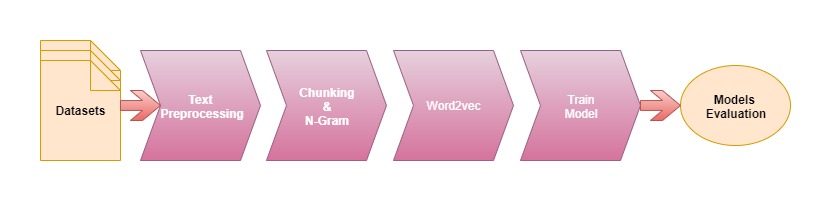}
\caption{Training Process of models} \label{Training process}
\end{figure}

We have trained our code review document corpus data sets with our chosen train models. Figure~\ref{Training process} shows the training process of our models

\section{Result \& Discussion}\label{Result & Discussion}
In this section, we explain the classification results, the evaluation contexts and the polarity estimation of our research. 
\subsection{Classification Results }\label{Clascsification_Results}
In order to classify the code review comments as Efficient or Not-Efficient or Some-How-Efficient or System-Generated, we used seven different classification algorithms. Table \ref{algo_accuracy} shows the accuracy of different classifiers we used.
\begin{table}[hbt!]
\centering
\setlength{\tabcolsep}{8pt}
\caption{Accuracy of different classifiers}\label{algo_accuracy}
\begin{tabular}{l c}
\hline
Algorithm & Accuracy (\%)\\
\hline
Na\"ive Bayes  & 81.32\\
Multi nomial Na\"ive Bayes (MNB) &  82.76 \\
Bernoulli Na\"ive Bayes  & 80.11\\
Logistic Regression  & 82.50\\
Stochastic Gradient Descent (SGD) & 78.55\\
Linear Support Vector Classifier(SVC) & 83.09\\
Nu support vector classifier & 79.75\\
\hline
\end{tabular}
\end{table}
It is terrible to see some of the widely used algorithms
fail to achieve satisfactory performance for this case.
Most notably, the Na\"ive Bayesian, Linear Support Vector Classifier(SVC), Multi nomial Na\"ive Bayes (MNB), Bernoulli Na\"ive Bayes and Logistic Regression classifiers achieve an average accuracy of about 81\%. However,
other variants of Stochastic Gradient Descent (SGD) and Nu support vector classifier  gave accuracy of more than 75\% but less than 80\% in some cases. The Linear Support Vector Classifier(SVC) classifier gives the best accuracy.

\subsection{Corpus Validation}
In our corpus datasets that we used to train the models, contains 13557 comments (Efficient-3326, Not-Efficient-419, Some-How-Efficient-975,and System-Generated-8837 lines comments). We train our models considering 75\% as training dataset of our code review corpus and test the models with remaining 25\% test dataset of corpus. Test dataset is used to provide an unbiased evaluation of a final model to fit with the training dataset. Section~\ref{Clascsification_Results} shows the details the validation results.

\subsection{Sentiment Estimation}
In this section, we explain the Sentiment results, the evaluation contexts and the polarity estimation of our research.
\paragraph{\textbf{Sentiment Consideration:}}
We present the words polarity in Table \ref{SentimentRatio}.
Our train model gives most informative features words in trained data sets. In the first instance, model gives informative words"classifier.show-most-informative-features()" base on our sentiment ratio score. 
\begin{table}[hbt!]
\centering
\setlength{\tabcolsep}{8pt}
\caption{Sentiment Ratio}\label{SentimentRatio}
\begin{tabular}{l c c}
\hline
Words & Sentiment & Sentiment Ratio\\\hline
Need & Efficient & 92.8\\
Please & Efficient & 98.5\\
Merged & System-Generated & 65.7\\
Ok & Not-Efficient & 78.5\\
Thanks & Not-Efficient & 85.6\\
Think & Some-How-Efficient & 82.7\\
\hline
\end{tabular}
\end{table}

\subsection{Precision, Recall and F-measure for Code Reviews Data sets}
As we found highest accuracy with Linear Support Vector Classifier(SVC), so we trained our model by Linear Support Vector Classifier(SVC). This time, we run random shuffle in our trained date sets (Efficient,Not-Efficient,Some-How-Efficient and System-Generated). This classifier gives the accuracy measuring the trained data set. We collect the reference values and observed values for each label (Efficient,Not-Efficient,Some-How-Efficient and System-Generated), then use those sets to calculate the precision, recall, and F-measure of the Linear Support Vector Classifier(SVC). Shows the resultant  confusion matrix is a table \ref{matrix} that is often used to describe the performance of a classification model on a set of test data for which the true values are known. It allows the visualization of the performance of an algorithm.
\begin{table}[hbt!]
\centering
\setlength{\tabcolsep}{8pt}
\caption{Confusion Matrix of code Reviews Data sets }\label{matrix}
\begin{tabular}{l c cc}
\hline
Sentiment & Precision & Recall & F-measure \\
\hline
Efficient & 0.67 & 0.81 & 0.73\\
Not-Efficient &  0.64 & 0.58 & 0.60\\
Some-How-Efficient &  0.62 & 0.68 & 0.64\\
System-Generated &  0.74 & 0.78 & 0.75\\
\hline
\end{tabular}
\end{table}

\subsection{Sentiment Polarity Considering Procedure}
In this section, train model gives sentiment against our code review comments. Instance model gives comments sentiment based on our training models.We will multiply the answer by 100\% so it's in a percentage and we will even denote this accuracy percent. Our models give comments sentiment the confidential score between 0 and 1.0 it gives the result as sentiment. Table \ref{confidential} shows some of the confidential score of code review comments.
\begin{table}[hbt!]
\centering
\setlength{\tabcolsep}{10pt}
\caption{Sentiment result of comments}\label{confidential}
\begin{tabular}{l c c}
\hline
Review comments & Sentiment & Confidential \\ &&score \\\hline
Please maintain same coding rules  & Efficient & 0.7 \\applied in this project. & & \\\hline
Sure I will change it. 
& Not-Efficient
& 0.6\\\hline
Should there be a new line? 
& Some-How-Efficient
& 0.7\\please check. & & \\\hline
Change has been successfully merged
& System-Generated
& 0.8\\by Rifat Hasan& & \\
\hline
\end{tabular}
\end{table}

\section{Conclusion}\label{Conclusion}
The review comments are the influential element of any bug issue of code. In this paper, we proposed a technique to identify the code review comments from  GitHub repositories projects in terms of semantic orientation using a machine learning approach. The objective is to find out the context based tagging of review comments to avoid prejudicing the coder. Our work will mostly help the programmers to make decision based on review comments by avoiding misconception . Moreover, this technique can also be implemented in a tool for identifying GitHub or Gerrit project repositories.

In the future, current work can be extended to analyze the review comments and developer reviews together to find out the most authenticate comments on GitHub.

%
%

\bibliographystyle{plain}
\bibliography{references}


\end{document}